**Understanding critical currents in superconducting cuprate tapes.**

Charles Simon[1*]

1 Centre National de la Recherche Scientifique (CNRS) and Université Grenoble Alpes, Laboratoire National des Champs Magnétiques Intenses (LNCMI), Grenoble, 38120 France.

**ABSTRACT**. One of the key challenges in the fabrication of superconducting coils using cuprate tapes is the parametrization of the critical currents and their dependence on magnetic field, temperature, and angle. Discussions at the Magnet Technology Conference (MT29) in Boston (2025) highlighted the need for standardized characterization and a better understanding of these tapes. Without a shared understanding of the physical phenomena governing critical currents, progress in this area remains difficult. We propose to analyze existing data using a model that explains most observed features. Although the model proposed by P. Mathieu and Y. Simon was published 20 years ago, it remains relatively unknown and certainly unused among engineers in the field, although many physicists were convinced of its validity — a consensus not reflected in the literature. The Mathieu/Simon (MS) model emphasizes the importance of surface pinning mechanisms, which dominate critical currents across a large part of the phase diagram of YBaCuO. Unlike strong and weak pinning mechanisms, which are commonly assumed to be dominant, the MS model accurately predicts the order of magnitude of experimentally measured values, suggesting it should at least be considered the dominant mechanism. The results of calculations based on this model are presented and compared with experimental data, offering directions for the development of new materials.

## I. INTRODUCTION.

One of the key limiting factors in the production of superconducting wires and tapes is the critical current density, particularly its rapid degradation with increasing temperature and magnetic field. State-of-the-art $YBa_2Cu_3O_{7-\delta}$ (YBCO) tapes are fabricated by depositing a few micrometers of YBCO onto substrates typically 50 to 100 micrometers thick. Established techniques include Rolling Assisted Biaxially Textured Substrates (RABiTS) on nickel substrates and Ion Beam Assisted Deposition (IBAD) on stainless steel or Hastelloy. Additional thin-film deposition methods, such as Pulsed Laser Deposition (PLD) and Metal Organic Chemical Vapor Deposition (MOCVD), are also used. Buffer layers—commonly composed of MgO, $SrTiO_3$, $Y_2O_3$, or $Al_2O_3$—are incorporated to improve crystallographic matching and prevent oxygen or cation diffusion.

Discussions at MT29 (2025) [1] confirmed cuprate tapes as the most promising materials for near-future superconducting coil applications, given the lack of reliable alternatives (Bi-based tapes are too soft and present insufficient mechanical properties).

While deposition techniques have become more reliable and efficient over the past 25 years, fundamental methods have remained largely unchanged. A 2002 review revealed a universal thickness dependence of critical currents across diverse sample types (Fig. 2 of [2]), suggesting that the specific nature of defects responsible for vortex pinning may not be the primary determinant of critical current behavior. Recently, Senatore et al. [3] published a rather comprehensive study providing valuable data, which will be discussed in this paper. This underscores the urgent need to deepen our understanding of the physical mechanisms governing critical currents in these materials.

Before presenting experimental findings, we outline the key features of the Mathieu/Simon (MS) model.

## II. PRESENTATION OF MS MODEL

The model (MS model) proposed by P. Mathieu and Y. Simon [4,5,6] is a model describing the critical current and the vortices in a type-II superconductors. It introduces quantities averaged over lengths much larger than the intervortex distance $a_0$, let us define $\theta c$ as the critical maximum angle at magnetic fields large compared to $H_{c1}$. The distance $a_0$ is directly derived from the magnetic field B with the quantity $\varphi_0$ (the flux quantum carried by a single vortex), $a_0$ is typically 10nm at 30T. This approach is analogous to that proposed by Bekarevich and Khalatnikov for vortex motion in superfluid helium (see [4]).

Bold characters represent vectors and brackets denote spatially averaged values, <B> for the magnetic field, <Js> for the superconducting current of the Cooper pairs, <N> for the vortex density, <ν> for the unit vector in the direction of the vortex density <ω>, m is the mass and e the charge of the electron respectively. The penetration of superconducting screening currents is governed by two fundamental equations derived from the London and Maxwell equations:

*Contact author: charles.simon@lncmi.cnrs.fr

$$\langle B\rangle - m/2e^2 \,\mathrm{curl}\, \langle Js\rangle = \langle\omega\rangle = \langle N\rangle\, \varphi_0\, \nu \qquad (1)$$

$$\mathrm{curl}\, \langle B\rangle = \mu_0 \langle Js\rangle \qquad (2)$$

For a slab of thickness t perpendicular to the magnetic field, the penetration depth of the averaged magnetic field is given by (see Fig. 1):

$$\mathrm{curl\,curl}\, \langle B\rangle = \mu_0\, \mathrm{curl}\, \langle J\rangle = 2\mu_0\, e^2/m \langle B\rangle + \mathrm{curl}\, (\langle N\rangle\, \varphi_0\, \nu) \qquad (3)$$

This is the standard situation in which the London penetration depth $\lambda_L$ appears in the absence of a magnetic field. ($\lambda_L = 1/\sqrt{2\mu_0\, e^2/m}$) $\lambda_{MS}$ varies from $\lambda_L$ at low field ($\langle N\rangle \sim 0$) to $a_0$ at high field. The formula can be approximated by

$$1/\lambda_{MS}^2 = 1/\lambda^2_L + 1/a^2_0 \qquad (4)$$

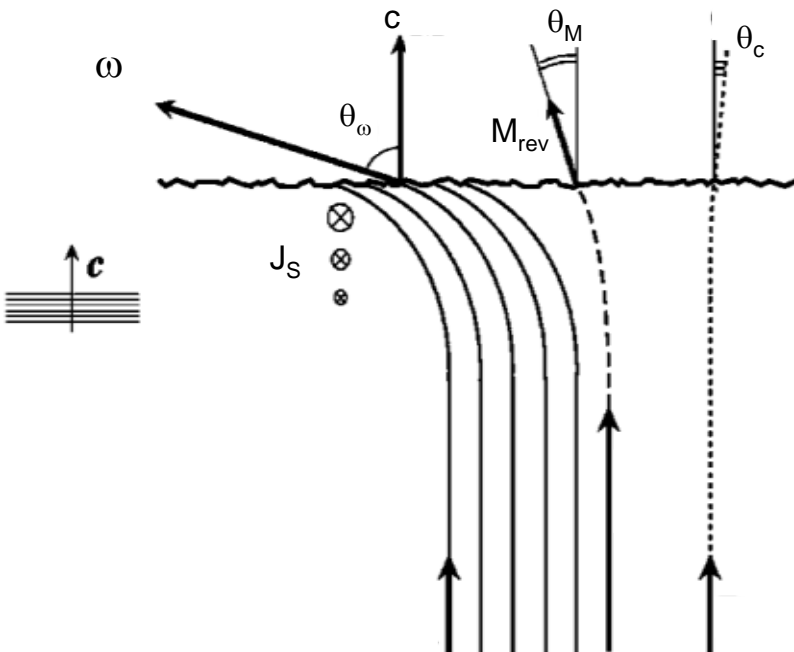

FIG. 1. Schematic situation of a finite surface roughness showing the definition of angles and quantities (adapted from [12]).

The critical step in solving these equations is introducing boundary conditions. While periodic boundary conditions in reciprocal space yield trivial solutions, realistic boundary conditions in real space are essential for the model. For this slab of thickness t (along z) and width w (along x) with two perfectly flat surfaces (infinite in the current direction y), vortices cross the surfaces perpendicularly, $\langle \mathbf{B}\rangle$ is uniform and $\langle \mathbf{Js}\rangle$ is exactly zero (from Eq. 2)). This situation (no critical current) is uninteresting for transport current. Conversely, the presence of a non-zero averaged superconducting current $\langle \mathbf{Js}\rangle$ (along y) means the vortices will bend within the thickness of the slab (from Eq.2, $\partial \langle B\rangle_x/\partial z - \partial \langle B\rangle_z/\partial x \neq 0$). If the surface has non-zero roughness, vortices may cross the average surface at a finite angle. Let name $\theta c$ the critical maximum angle for this rough surface, allowing vortices to bend at a critical maximum angle. The critical current density is then given by:

$$\mu_0 \langle Js\rangle /w = \langle M_{rev}\rangle(B) \sin\theta c \qquad (5)$$

where $M_{rev}(B)$ is the conjugated quantity of B in the free energy of the system, denoted $\varepsilon(B)$ is MS papers.

In order to estimate the free energy of the mixed state, good approximations of $M_{rev}(B)$ can be found in the literature (see for example H. Brandt analysis of this problem) [7]

London model valid in a regime between $B_{c1}$ and $B_{c2}$

$$M_{rev}(B) = \varphi_0/(8\pi\, \mu_0\, \lambda_L^2)\, \ln(\omega/\mu_0 B_{c2}) \qquad (6)$$

Or Abrikosov model valid close to $B_{c2}$ [8]

$$M_{rev}(B) = (B_{c2} - \omega/\mu_0)/(1.16\,(2K^2 - 1) + 1) \qquad (7)$$

Additionally,

$$M_{rev}(B=0) = B_{c1} \qquad (8)$$

A useful ad hoc interpolation formula was proposed by Plaçais et al [9] :

$$M_{rev}(B) = (B_{c2} - \omega/\mu_0)\,(\ln(1.68K) + \tfrac{1}{2}\ln((\mu_0 B_{c2} + 7\omega)/(\mu_0 B_{c2} + 4K^2\omega))) \qquad (9)$$

This interpolation formula was tested for YBaCuO in the A. Pautrat thesis [10] by comparing this approximation to the results of the numerical calculations made by Hao and Clem [11]. For YBaCuO, the presence of an anisotropy $\gamma$ (about 5 to 7 depending on the exact doping) complicates the model. A modified MS theory for anisotropic systems [12] accounts for this anisotropy in $NbSe_2$, where $M_{rev}$ is a vector not parallel to B but at an angle $\theta_M$ (defined from the surface normal). The angle $\theta_\omega$ between **B** and the c-axis (perpendicular to the flat surface) is related to $\theta_M$ by

$$\tan(\theta_\omega) = \gamma \tan(\theta_M) \qquad (10)$$

Assuming that the angle $\theta_\omega$ remains small, $\theta_\omega$ can be approached by $\gamma\, \theta_M$ so Eq. (5) becomes:

$$\mu_0 \langle Js\rangle/w = M_{rev}(B)\, \tan\theta c / \gamma \qquad (11)$$

The critical current does not depend on the sample thickness as long as the thickness is larger than the critical current penetration depth $\lambda_{MS}$. The saturated critical current for infinite thickness is given by (11), without any adjustable parameters if the surface roughness $\theta c$ is measured.

$$Ic(B)/w = M_{rev}(B) \tan\theta c / \gamma \qquad (12)$$

where $M_{rev}(B)$ will be replaced by eq. 9.

## III. COMPARISON WITH EXISTING DATA AND PREDICTIONS FOR FUTURE IMPROVEMENTS OF MATERIAL PERFORMANCES

### A. Surface critical current Ic(0)/w with zero applied magnetic field : thickness dependence

Equation 12 predicts that the critical current does not depend on the sample thickness t. Data available in the literature are rather limited, but Ref. [2] provides thickness dependence for cuprate tapes. We extracted data for IBAD samples on MgO (Fig. 2 of [2]) and replotted Ic/w vs. thickness. A clear saturation above 2 µm is observed, as expected. The critical thickness corresponds to the critical current penetration depth $\lambda_{MS}$, and the experimental value of 2 µm is reasonable for $\lambda_{MS}$ at 75 K and low magnetic field. The saturated Ic/w (300 A/cm-width) agrees well with $B_{c1}$=250 A/cm, $\gamma$=7, and measured surface roughness $\theta c$=10° (Eq. 12).

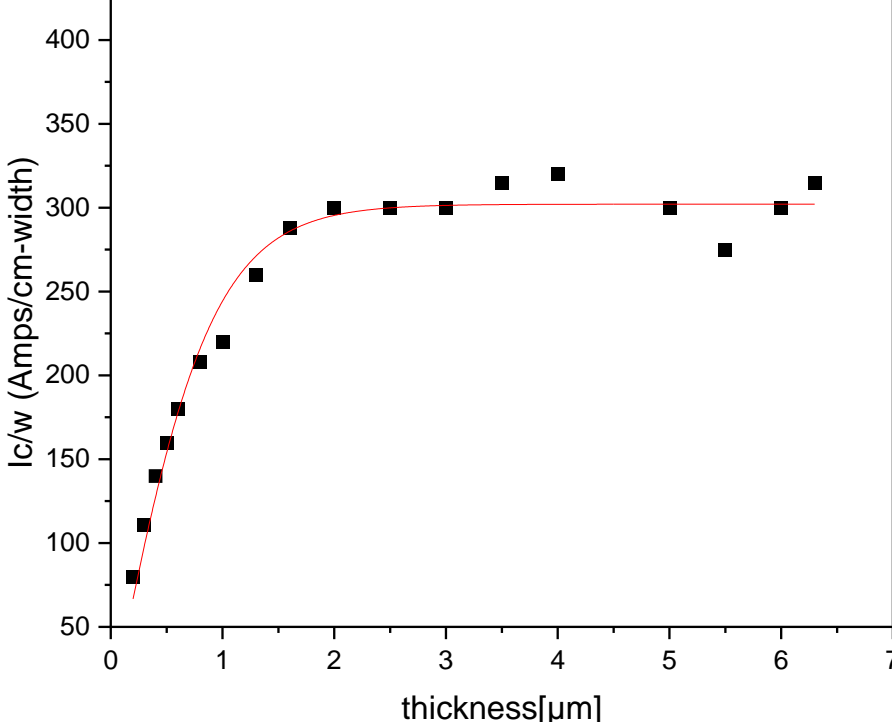

FIG. 2. Comparison of the data from ref [2] (black points) with the MS model (red line) for YBaCuO IBAD samples on MgO at 75K and zero magnetic field.

This type of comparison is rather rare in the literature. In most of the studies, the type of surface roughness depends strongly on the thickness and no easy regularity is observed.

In addition, the literature often discusses the origin of vortex pinning in YBaCuO (strong vs. weak pinning), but the surface pinning effect is typically dismissed without quantitative evaluation. The very comprehensive Blatter et al. publication [13] provides a lot of models for strong and weak pinning which can be used for analysis, but most of them are not really quantitative in reality. In contrast, I will show that the MS model provides directly the correct order of magnitude without invoking bulk pinning with parameters of the right order of magnitude.

### B. Surface critical current Ic(B)/w : magnetic field dependence and irreversibility line

Screening currents are of the order of $B_{c1}$, so the difference between the applied field and the field inside the sample will be less than $B_{c1}$. In the case of YBaCuO, $B_{c1}$ and of $B_{c2}$ are very different, (by the factor $B_{c2}/B_{c1}$ =100), this difference can be neglected above 1T.

YBaCuO and many cuprates exhibit an irreversibility line, where the critical current vanishes at $B_{irr}$ well below $B_{c2}$. This line is often attributed to vortex lattice melting, but such conclusions are limited to high temperatures and low fields (above 74 K, below 10 T), where thermal fluctuations dominate [14]. At lower temperatures and higher fields, the situation is less clear. For example, neutron diffraction shows the vortex lattice persists above the "irreversibility line" in Bi-2212 [15]. Here, we will focus on temperatures below 74 K, where thermal fluctuations are negligible.

Lazard et al. [12] applied the MS model to anisotropic $NbSe_2$ ($\gamma$=3). A key result is the appearance of a characteristic field $B_{irr}$, where Is/w reaches zero well below $B_{c2}$, as observed experimentally in one of the few measurements at fields higher than 20T [16]:

$$\mu_0 I_s/w = B_{c2} / (2 \beta_A K^2) \tan\theta_c \left((1+\gamma^2 \tan^2\theta_c)^{-1/2} - B/B_{c2}\right) \qquad (13)$$

By introducing

$$B_{irr} = B_{C2}(1+\gamma^2 \tan^2\theta_c)^{-1/2} \qquad (14)$$

One gets

$\mu_0 I_s/w = (B_{irr} - B) \tan \theta_c / (2 \beta_A K^2) = 5\ 10^{-4} (B_{irr} - B)$ (15)

$I_s/w$ (Amps/cm-width) = $5 (B_{irr} - B)$ (Tesla) and $B_{irr}/B_{c2} = 0.82$.

The critical current $I_s/w$ decreases linearly and vanishes at $B_{irr}$. In the case of YBaCuO, the anisotropy $\gamma$ is 7 and the K value is about 100, $\beta_A = 1.16$ for the triangular lattice. $\tan\theta_c$ is the surface roughness assumed to be 0.11, this explains the irreversibility line. For Bi-2212 ($\gamma \approx 200$), this agreement is reinforced, though precise calculations are more complex.

Between $B_{irr}$ and $B_{c2}$, vortices bend near the surface, creating a normal-state layer that screens the surface pinning mechanism.

To go beyond this approximation (B much larger than $B_{c1}$, incorporating Eq. (9) into Eq. (12) gives:

$$Ic (B) /w = \tan\theta c / (2 \mu_0 \beta_A K^2) (B_{irr} - B) (\ln(1.68K) + ½ \ln(B_{irr}+7B) - \ln(B_{irr} + 4K^2B)) \quad (16)$$

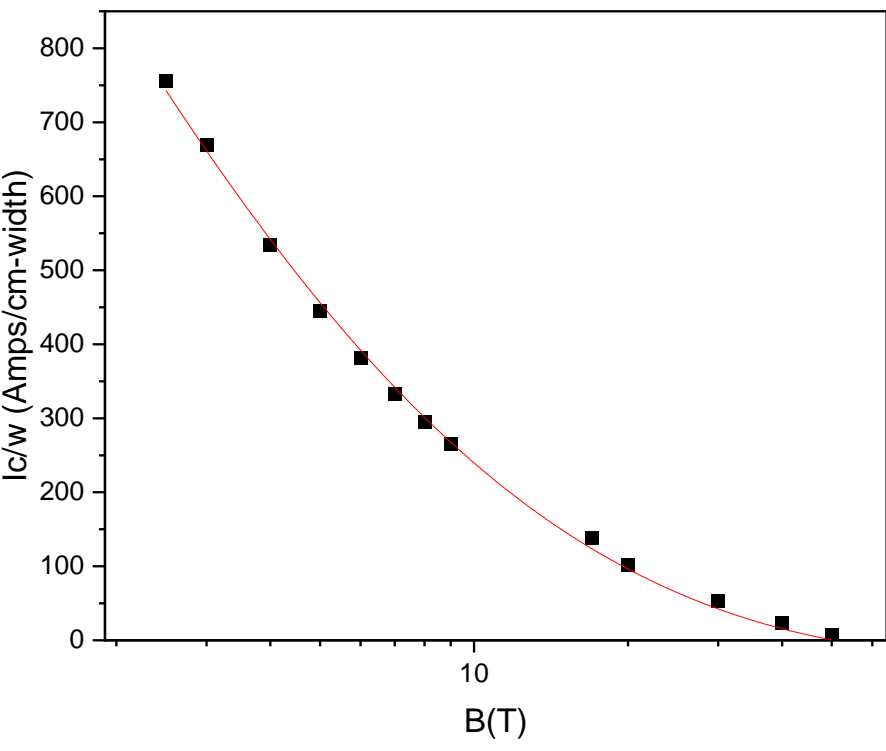

FIG. 3. Comparison between experimental data at 27K extracted from [14] and the MS model using the Plaçais interpolation formula of equation 16. The fitting parameters are: $B_{irr}$(27K) = 50.6T, tanθc =0.11, $\gamma$ =7, K=101.1.

This model accurately predicts the magnetic field dependence of the critical current in this range of magnetic fields (Fig. 3). The three adjustable parameters take very reasonable values for YBCuO at 27K : $B_{irr}$(27K) = 50.6T, tanθc =0.11, $\gamma$ =7, K=101.1. It is very remarkable that this model provides a good understanding of the vortex physics in this range of temperature and magnetic field.

Far from $B_{c2}$, the logarithmic approximation for $M_{rev}$ (Eq. 6) fits experimental data well (Fig. 4).

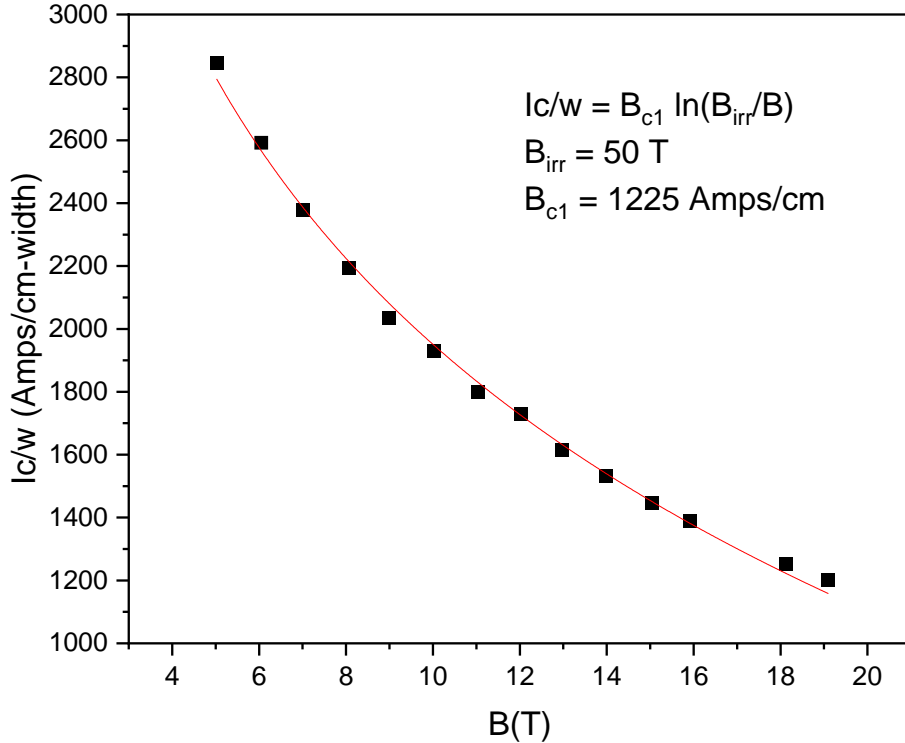

FIG. 4. Data from ref [3] fitted by equation 12 with $M_{rev}$ approximated by the London model of Eq. 6 (logarithmic approximation).

This logarithmic dependence is directly related to the magnetic field dependence of vortex-vortex interaction in this range.

## IV. CONCLUSIONS

The MS model accurately predicts the order of magnitude and the magnetic field dependence of the critical current, including the position of the irreversibility line. We propose using this interpolation formula to systematically study the magnetic field dependence of critical current in YBaCuO tapes. Also, at low temperature and high fields, tapes with a much thinner (~15-30 nm) layer of cuprates than what is typical (~1um) should retain exactly the same critical current. This discussion guides sample thickness selection. Per Eq. (4) at 4 K and 40 T (typical for magnets) $\lambda_{MS}$ ~7 nm is enough to saturate the critical current; even at 10 T, 14 nm is adequate. This prediction is a major conclusion of this work: **critical current flows only near the sample surfaces. Hence, in moderate to high field applications, most of the thickness of the 1µm-thick cuprate layer in a typical REBCO tape, does not actually bearing any critical current.** Stated otherwise, the bulk of the cuprate layer could be removed while still maintaining the same low-temperature high-field critical current.

## ACKNOWLEDGMENTS

Charles Simon acknowledges stimulating discussions with Thierry Klein, Maxime Leroux, Johan Graglia and Xavier Chaud.